\newcommand{\AaA}{A\&A}
\newcommand{\ApJ}{ApJ}
\newcommand{\MNRAS}{MNRAS}
\newcommand{\Natur}{Nature}
\newcommand{\PhLB}{Phys. Lett. B}
\newcommand{\PhL}{Phys. Lett.}
\newcommand{\PhRvD}{Phys. Rev. D}
\newcommand{\PhRvL}{Phys. Rev. Lett.}

\documentclass{aa}
\usepackage{graphics}
\usepackage{amsfonts}
\begin{document}
\def\gtsima{$\; \buildrel > \over \sim \;$}
\def\simgt{\lower.5ex\hbox{\gtsima}}
\thesaurus{03.13.2, 03.13.6, 03.20.1}
\title{Searching for non-gaussianity: Statistical tests}
\author{O. Forni \and N. Aghanim}
\offprints{N. Aghanim, aghanim@ias.fr}
\institute{IAS-CNRS, Universit\'e Paris Sud, B\^atiment 121, F-91405 Orsay 
Cedex}
\date{Received date / accepted date}
%\titilerunning
%\titilerunning
\maketitle
\markboth{Searching for non-gaussianity: Statistical tests}{}
\begin{abstract}
Non-gaussianity represents the statistical signature of physical processes 
such as turbulence. It can also be used as a powerful tool to discriminate 
between 
competing cosmological scenarios. A canonical analysis of non-gaussianity is
based on the study of the distribution of the signal in the real (or direct)
space (e.g. brightness, temperature).\par
This work presents an image processing method in which we propose statistical
tests to indicate and quantify the non-gaussian nature of a signal. Our
method is based on a wavelet analysis of a signal. Because the temperature
or brightness distribution is a rather weak discriminator, the search for the 
statistical signature of non-gaussianity relies on the study of the coefficient
distribution of an image in the wavelet decomposition basis which is much
more sensitive.\par
We develop two statistical tests for non-gaussianity. In order to
test their reliability, we apply them to sets of test maps
representing a combination of gaussian and non-gaussian signals.
We deliberately choose a signal with a weak non-gaussian signature and we find 
that such a non-gaussian signature is easily detected using our 
statistical discriminators.
In a second paper, we apply the tests in a cosmological context.
\end{abstract}
\section{Introduction}
 Non-gaussianity is a very promising way of characterising some important 
physical 
processes and has many applications in astrophysics. In fluid mechanics, the
non-gaussian signature of the probability density functions of velocity is 
used as
evidence for turbulence (e.g. Chabaud et al. (1994)). Astrophysical fluids with 
very large Reynolds numbers, such as the interstellar medium, are expected to 
be turbulent. If it exists, turbulence should play a leading role in the
triggering of star formation, in determining the dynamical structure of the
interstellar medium and its chemical evolution. Non-gaussianity is a
tool for indicating turbulence. Indeed, recent studies on the non-gaussian
shape of the molecular line profiles can be interpreted as evidence for 
turbulence (Falgarone \& Phillips 1990; Falgarone et al. 1994; Falgarone \&
Puget 1995; Miville-Desch\^enes et al. 1999).
Non-gaussianity is also used as the indicator of coronal heating due to
magneto-hydro-dynamical turbulence (Bocchialini et al. 1997; Georgoulis et al.
1998).
Within the cosmological framework, the statistical nature of the Cosmic 
Microwave Background (CMB) temperature, or brightness anisotropies, probes the
origin of the initial density perturbations which gave rise to cosmic 
structures (galaxies, galaxy clusters). The inflationary models 
(Guth 1981; Linde 1982) predict gaussian distributed density perturbations,
whereas the topological defect models 
(Vilenkin 1985; Bouchet 1988; Stebbins 1988; Turok 1989) generate a non-gaussian 
distribution. Because the nature of the initial density perturbations is a 
major question in cosmology, a lot of statistical tools have been developed 
to test for non-gaussianity. \par
In order to test for non-gaussianity, one can use traditional methods 
based on the distribution of temperature anisotropies. The simplest tests
are based on the third and/or fourth order moments (skewness and kurtosis) of 
the 
distribution, both equal to zero for a gaussian distribution. Another test for 
non-gaussianity through the temperature distribution is based on the study of 
the cumulants (Ferreira et al 1997; Winitzki 1998). The n-point 
correlation functions also give very valuable statistical information.
In particular, the three-point function, and 
its spherical harmonic transform (the bispectrum), are appropriate tools for 
the detection of non-gaussianity 
(Luo \& Schramm 1993; Kogut et al. 1996; Ferreira \& Magueijo 
1997; Ferreira et al. 1998; Heavens 1998; Spergel \& Goldberg 1998). An
investigation of the detailed behaviour of each multipole of the CMB angular
power
spectrum (transform of the two-point function) is another non-gaussianity
indicator (Magueijo 1995). Finally, other tests of non-gaussianity are 
based on topological pattern statistics (Coles 1988; Gott et al. 1990). 
\par\medskip
The works of Ferreira \& Magueijo (1997) on the search for 
non-gaussianity in Fourier space, and of Ferreira et al. (1997) on the
cumulants of wavelet transformed maps, have shown that these approaches 
allow the study of characteristic scales, which is particularly interesting 
when studying a combination of gaussian and non-gaussian signals as a 
function of scale. In the present work, we study non-gaussianity in a
wavelet decomposition framework, that is, using the coefficients in the wavelet
decomposition. We 
decompose the image of the studied signal into a wavelet basis and analyse the 
statistical properties of the coefficient distribution. We then
search for reliable statistical diagnoses to distinguish between gaussian and
non-gaussian signals. 
The aim of this study is to find suitable tools for demonstrating and to
quantifying the non-gaussian signature of a signal when it is combined with a 
gaussian distributed signal of similar, or higher, amplitude. \\
In section 2, we describe the methods for wavelet decomposition and 
filtering. We then
briefly describe the characteristics of the wavelet we use in our study. We 
also give the main characteristics of the test data sets
used in our work. In section 3, we present the statistical criteria we 
developed to test for non-gaussianity. We then apply the tests to sets of 
gaussian test maps (Sec. 4). We also apply them, in section 5, to sets of 
non-gaussian maps as well as combinations of gaussian and non-gaussian signals
with the same power spectrum. In section 6, we present and apply the
detection strategy we propose to quantify the detectability of the non-gaussian
signature. Finally, in section 7 we conclude and present our main results.
\section{Analysis}
\subsection{The wavelet decomposition}\label{sec:wdec}
Wavelet transforms have been widely investigated recently because of their 
suitability for a large number of signal and image processing tasks. 
Wavelet analysis involves a convolution of the signal with a convolving
function or a kernel (the wavelet) of specific mathematical properties. 
When satisfied, these properties define an orthogonal basis, which 
conserves energy, and guarantee the existence of an inverse to the wavelet 
transform.\par
The principle behind the wavelet 
transform, as described by Grossman \& Morlet (1984), Daubechies (1988) and 
Mallat (1989), is to hierarchically decompose an input signal into a series 
of successively lower resolution reference signals and their associated detail 
signals. At each decomposition level, L, the reference signal has a 
resolution reduced by a
factor of $\mathrm{2^L}$ with respect to the original signal.
Together with its respective detail signal, each scale contains the 
information needed to reconstruct the reference signal at the next higher
resolution level.
Wavelet analysis can therefore be considered as a series of bandpass filters
and be viewed as the decomposition of the signal into a set of independent, 
spatially oriented frequency channels. Using the 
orthogonality properties, a function in this decomposition can be completely
characterised by the wavelet basis and the wavelet coefficients of the
decomposition. \par
 The multilevel wavelet transform (analysis stage) decomposes the signal into 
sets of different 
frequency localisations. It is performed by iterative application of a pair of
Quadrature Mirror Filters (QMF). A scaling
function and a wavelet function are associated with this analysis filter bank. 
The continuous scaling 
function $\phi_{A}(x)$ satisfies the following two-scale equation:
\begin{equation}
\phi_{A}(x)= \sqrt{2} \sum_n h_{0}(n)\phi_{A}(2x-n),
\end{equation}
where $\mathrm{h_{0}}$ is the low-pass QMF.
The continuous wavelet $\psi_{A}(x)$ is defined in terms of the scaling 
function and the high-pass QMF $\mathrm{h_{1}}$ through:
\begin{equation}
\psi_{A}(x)=\sqrt{2} \sum_n h_{1}(n)\phi_{A}(2x-n).
\end{equation}

The same relations apply for the inverse transform (synthesis stage) but,
generally, different scaling function and wavelet ($\phi_{S}(x)$ and 
$\psi_{S}(x)$) are associated with this stage:
\begin{equation}
\phi_{S}(x)= \sqrt{2}\sum_n g_{0}(n)\phi_{S}(2x-n),
\end{equation}
\begin{equation}
\psi_{S}(x)= \sqrt{2}\sum_n g_{1}(n)\phi_{S}(2x-n).
\end{equation}

Equations (1) and (3) converge to compactly supported basis functions when 
\begin{equation}
\sum_n h_{0}(n)=\sum_n g_{0}(n)=\sqrt{2}.
\end{equation}

The system is said to be biorthogonal if the following conditions are 
satisfied:
\begin{eqnarray}
& & \int_{\mathbb{R}}\phi_{A}(x)\phi_{S}(x-k)dx=\delta(k) \\
& & \int_{\mathbb{R}}\phi_{A}(x)\psi_{S}(x-k)dx=0 \\
& & \int_{\mathbb{R}}\phi_{S}(x)\psi_{A}(x-k)dx=0 
\end{eqnarray}

Cohen et al. (1990) and Vetterli \& Herley (1992) give a 
complete treatment of the relationship between the filter coefficients and the 
scaling functions.
\par
\begin{figure}
%\hbox{\epsffile{ds8533f1.eps}}
\resizebox{12cm}{!}{\includegraphics{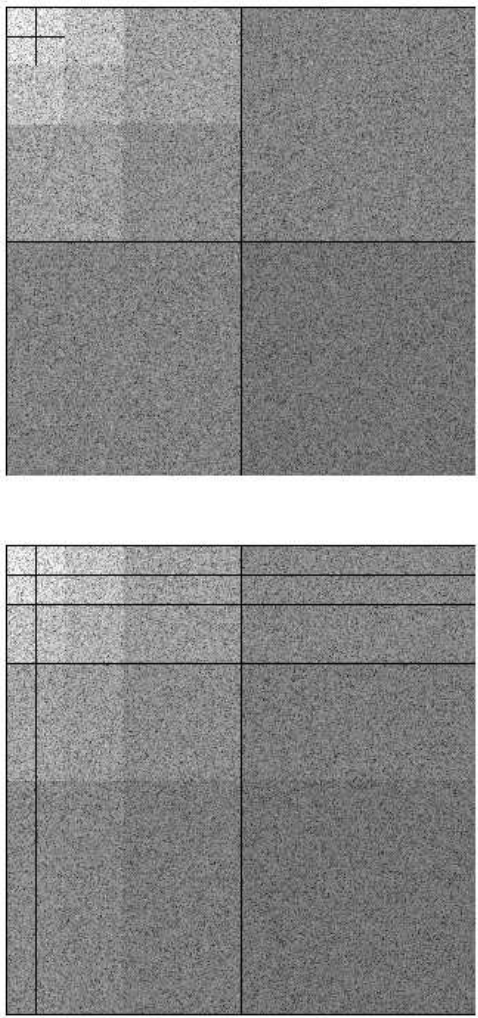}}
\caption{{\small\it In the upper panel, a 4 level 
 ``dyadic'' decomposition is used, in the lower, a 4 level ``pyramidal''
 decomposition is used. In both cases, the diagonal squares and the last 
decomposition level are identical and the grey levels indicate the amplitude of
the wavelet coefficients.}}
\label{fig:dpwvl}
\end{figure}
When applied to bi-dimensional data (typically images), three main types of
decomposition can be considered: dyadic (or octave band), pyramidal and 
uniform. 
\begin{enumerate}
\item
A ``dyadic'' decomposition refers to a transform in which only the reference
sub-band (low-pass part of the signal) is decomposed at each level. In this 
case, the analysis stage is 
applied in both directions of the image at each decomposition level. The total 
number of sub-bands after L levels of decomposition is then 3L+1 (Fig. 
\ref{fig:dpwvl}, upper panel).
\item
A ``pyramidal'' decomposition is similar to a ``dyadic'' decomposition in the 
sense that only the reference sub-band is decomposed at each level, but it
refers here to a transform that is performed separately in the two directions
of the image. The total number of sub-bands after L levels of 
decomposition is then $\mathrm{(L+1)^2}$ (Fig. \ref{fig:dpwvl}, lower 
panel).
\item
A ``uniform'' decomposition refers to one in which all sub-bands are 
transformed at each level. The total number of sub-bands after L levels of 
decomposition is then $\mathrm{4^L}$.
\end{enumerate}
\par
The wavelet functions are localised in space and, simultaneously, they
are also localised in frequency. Therefore, this
approach is an elegant and powerful tool for image
analysis because the features of interest in an image are present at 
different characteristic scales.  
Moreover, if the input field is gaussian distributed, the output is 
distributed the same way, regardless of the power spectrum. This arises from 
the linear transformation properties of gaussian variables. The distribution 
of the wavelet coefficients of a gaussian process is thus a gaussian.
Conversely, we expect that any non-gaussian signal will exhibit a 
non-gaussian distribution of its wavelet coefficients.
\par
\begin{figure}
%\hbox{\epsffile{ds8533f2.eps}}
\resizebox{10cm}{!}{\includegraphics{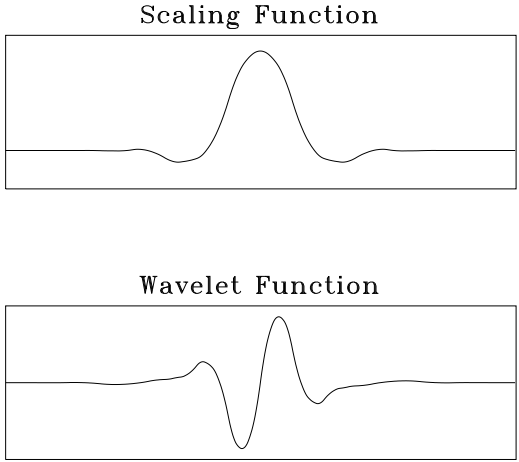}}
\caption{{\small\it Scaling function (top) and wavelet function (bottom) 
corresponding to the filter \#3 of Villasenor et al. (1995). Note that the
wavelet function is antisymmetric.}}
\label{fig:swfl}
\end{figure}

In our study, we have used bi-orthogonal wavelets, which are mainly used in
data compression, because of their better performance than
orthogonal wavelets, in compacting the energy into fewer significant 
coefficients.
There exist bi-orthogonal wavelet bases of compact support that are symmetric
or antisymmetric. Antisymmetric wavelets are proportional to, or almost
proportional to, 
a first derivative operator (e.g. the 2/6 tap filter (filter \#5) of 
Villasenor et al. (1995), or the famous Haar transform which is an orthogonal 
wavelet). Symmetric wavelets are proportional to, or almost proportional to, 
a second 
derivative operator (e.g. the 9/3 tap filter of Antonini et al. (1992)).
In the frame of detecting non-gaussian signatures, the choice of the wavelet
basis is critical because non gaussian features exhibit point sources or
step edges the wavelet must have a very good impulse response and a low shift 
variance, {\it i.e.} they better preserve the amplitude and the location of 
the details. Villasenor et al. (1995) have tested a set of bi-orthogonal filter
banks, within this context, to determine the best ones for image compression.
They conclude that even length filters have significantly less
shift variance than odd length filters, and that their performance in term of
impulse response is superior. In these filters, the high pass QMF is
antisymmetric which is also a desirable property in the sense that we will
also be interested in the statistical properties of the multi-scale gradients.
Consequently, in our study, we have chosen the 6/10 tap filter (filter \#3)
of Villasenor et al. (1995) (Fig. \ref{fig:swfl}) which represents the best 
compromise between all the criteria and energy compaction. 
Using this filter, we have chosen to
perform a four level dyadic decomposition of our data.
This particular wavelet and decomposition method have already been used for
source detection by Aghanim et al. (1998). 
With such a decomposition, we also
benefit from correlations between the two directions at each level, which
is not the case with the pyramidal decomposition that treats both directions as
if they were independent. Another advantage of this transform is that,
for each level of decomposition, or scale, we benefit from the maximum number 
of coefficients possible, which is crucial for the statistics.
\subsection{The test maps}
We use a test case which consists of sets of 100 maps of a gaussian signal 
and 200 maps of a non-gaussian signal, all having the same bell-shaped power 
spectrum and $512\times512$ pixels. One of the non-gaussian sets (100 maps)
consists of a distribution of disks with uniform
amplitude (top-hat profiles), generating step edges. The disks have different
sizes and amplitudes and are randomly distributed in the map. The signal is 
weakly skewed because the average skewness, which is the third moment of 
the distribution, computed over the 100 non-gaussian realisations is 
$\mu_3=-0.10\pm0.05$ (one sigma error for one realisation). Whereas the excess 
of kurtosis, the fourth moment ($\mu_4$) of the distribution, is 
$k=\mu_4/\mu_2^2 -3=1.09\pm0.17$. 
The second set of 100 non-gaussian maps consists of a distribution of gaussian 
profiles with different sizes and amplitudes. The skewness and excess of 
kurtosis of the 100 
statistical realisations of the non-gaussian distribution of gaussian profiles
are respectively $-0.06\pm 0.04$ and $1.19\pm 0.13$. The same 
quantities computed over the 100 gaussian maps are respectively 
$\mu_3=0.14\,10^{-2}\pm 2.15\,10^{-2}$ and $0.03\,10^{-2}\pm3.87\,10^{-2}$. 
These numbers should be equal to zero for a gaussian distribution. They are 
not because there is a statistical dispersion over a finite set of 
realisations. The purpose of this paper 
being to develop suitable statistical tests for non-gaussianity, we will not 
study other effects such as noise or beam dilution. These effects will be 
considered in an application of the method to the CMB signal in a second paper 
(Aghanim \& Forni 1999).
\section{Tests of non-gaussianity}
The most direct and obvious way of analysing the statistical properties of an 
image is to use the distribution of the pixel brightnesses, or temperatures,
together with the skewness and kurtosis. If the two quantities are different
from zero, they indicate that the signal is non-gaussian. However, a weak 
non-gaussian signal will 
hardly be detected through the moments of the temperature distribution.
Another way of addressing the problem is
to use the coefficients in the wavelet decomposition and to study their 
statistical properties which in turn characterise the signal. In 
fact, the wavelet coefficients are quite sensitive to variations (even weak 
ones) in the signal,
temperature or brightness, and hence to the statistical properties of the
underlying process. We have developed two tests which exhibit the 
non-gaussian characteristics of a signal using the wavelet coefficients.
Since our test maps are not skewed in the following  
we focus only on the results obtained using the fourth moment.\par
For the first discriminator, we study the statistical properties of the
distribution of the multi-scale gradient coefficients.
This method is appropriate when dealing with a non-gaussian process
characterised by sharp edges and consequently by strong gradients in the 
signal. 
Indeed, in any region where the analysed function is smooth the wavelet
coefficients are small. On the contrary, any abrupt change in the behaviour of
the function increases the amplitude of the coefficients around the 
singularity.
The detection of non-gaussianity is thus based on the search of these
gradients. In the dyadic wavelet decomposition one can discriminate between
the coefficients associated with vertical and horizontal gradients and the 
other coefficients. In our case, the vertical and horizontal gradients are
analogous to the partial derivatives, $\partial/\partial x$ and 
$\partial/\partial y$, of the signal.
Mallat \& Zhong (1992) give
a thorough treatment of the characterisation of signals from multi-scale edges.
We compute the quadratic sum of the coefficients, the quantity 
${\mathcal{G}}_L=\left(\partial/\partial x\right)_L^2 +\left(\partial/\partial y\right)_L^2$,
at each decomposition level $L$. This quantity represents the squared 
amplitude of the
multi-scale gradient of the image. In the following, we will however refer to
it as the multi-scale gradient coefficient. \par
The second statistical discriminator is based on the study of the wavelet
coefficients related to the horizontal, vertical and diagonal gradients. 
These coefficients are associated with the 
partial derivatives $\partial/\partial y$ and $\partial/\partial x$, as in 
the multi-scale gradient method, as well as with the cross derivative 
$\partial^2/\partial x\partial y$. The coefficients are computed at each
decomposition level and their excess of kurtosis with respect to a Gauss
distribution exhibits the non-gaussian signature of the studied signal. In
this context, the wavelet coefficients associated with the first derivatives
are obviously closely related to the multi-scale gradient. \\
In the following, we first apply our two tests to purely gaussian and 
non-gaussian maps. We then test the detectability of a non-gaussian
signal added to a gaussian one with same power spectrum and with 
increasing mixing ratios.
\section{Characterisation of gaussian signals}
\subsection{The multi-scale gradient and its distribution}
For the 100 gaussian maps, we find that the histogram of the 
multi-scale gradient coefficients can be fitted by the positive wing of the 
Laplace probability distribution function:
\begin{equation}
{\mathcal{H}}({\mathcal{G}}_L)=\frac{1}{\sqrt{2}\,\sigma}\,
\exp\left(-\frac{\sqrt{2}\,({\mathcal{G}_L}-\mu_1)}{\sigma}\right),
\end{equation}
where $\mu_1$ is the mean of the distribution (theoretically equal to zero) 
and $\sigma^2$ is its second moment. 

\begin{figure*}
%\epsfxsize=\columnwidth
%\hbox{\epsffile{ds8533f3.eps}}
\resizebox{\hsize}{!}{\includegraphics{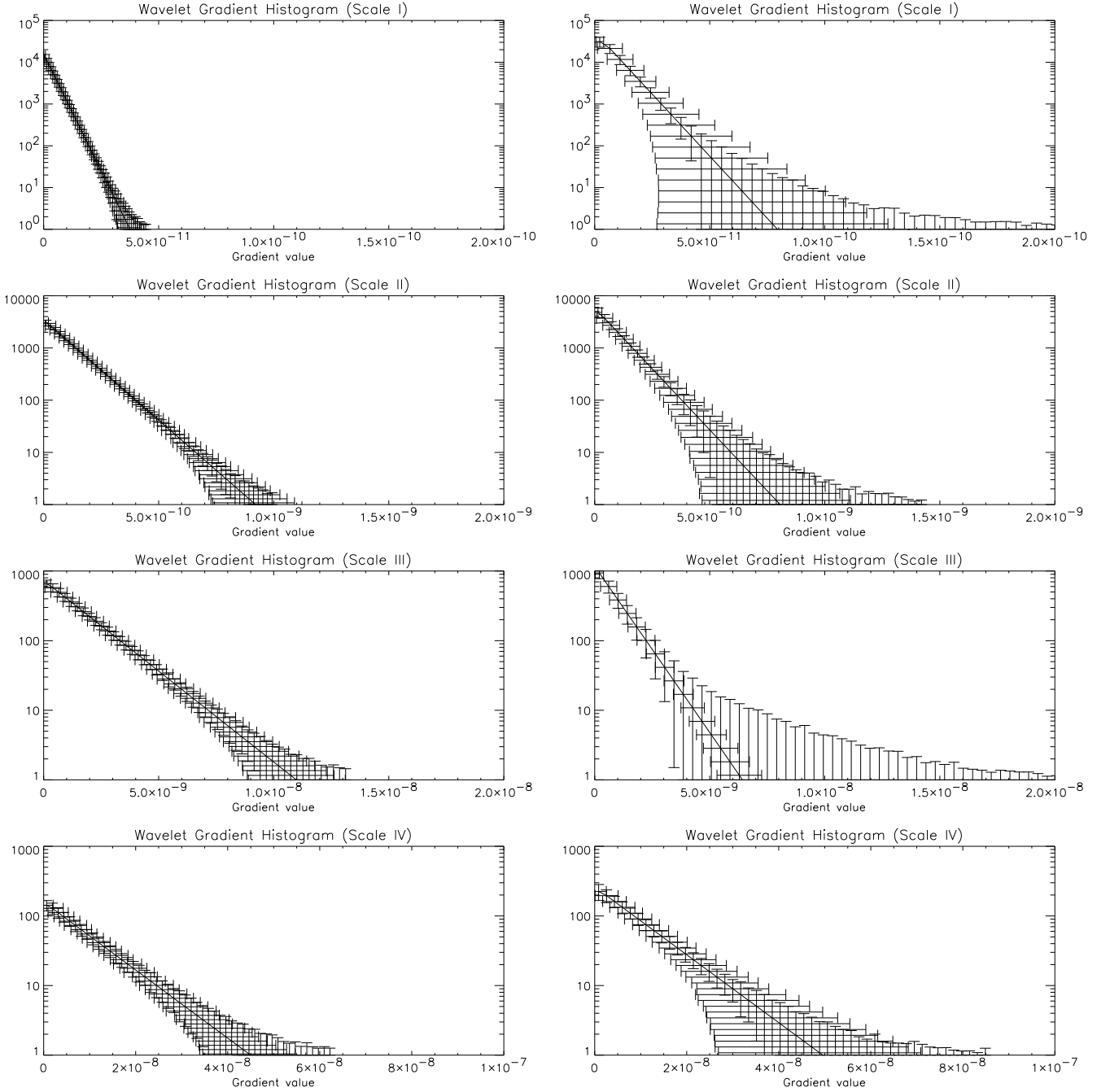}}
\caption{{\small\it Fits of the multi-scale gradient coefficient histograms 
obtained with the 100 statistical realisations. In all plots, the error bar 
is for one realisation. The 
left panels are given for the gaussian process whereas the right panels 
represent the 
non-gaussian signal. The wings at large multi-scale gradient indicate
the non-gaussian signature.}}
\label{fig:hist-gng}
\end{figure*}
\par
We plot in figure \ref{fig:hist-gng} (left panels) the distribution of the 
multi-scale gradient coefficients in the four decomposition scales for the
gaussian signal. For reasons of legibility, we have just
plotted the fit obtained with the 100 maps. The error bars represent a 
confidence interval (for one realisation) and account for the 
statistical dispersion of the realisations.\par
We can analyse the multi-scale
gradient distribution through its $n$th-order moments ($\mu_n$). In 
particular, we compute the excess of kurtosis using the second and 
fourth moments of the distribution($\mu_2$ and $\mu_4$). For a gaussian 
distribution, the normalised excess  
is zero. For a Laplace distribution, the fourth 
moment is given by $\mu_4=6\mu_2^2$. The normalised excess of kurtosis 
$k=\mu_4/\mu_2^2-6$ highlights the non-gaussian signature of a signal through 
the departure of the multi-scale gradient from a Laplace distribution. 
\par
At each decomposition level, we compute the normalised excess of kurtosis of 
the multi-scale gradient coefficients for the 100 gaussian maps, and we derive
a representative value of the distribution that is the mean $\overline k$ 
which we quote in Table 
\ref{tab:gng}. The results show that $\overline k$ 
is very close to zero. The $\sigma$ values correspond to the root mean square 
values with respect to the mean $\overline k$. The 
$\sigma$ values define a confidence interval, or a probability distribution
of the excess of kurtosis.
For the gaussian signal the upper and lower boundaries of this interval are 
equal suggesting that the $k$ values are gaussian distributed. The increasing 
$\sigma$ (Fig. \ref{fig:hist-gng}, left panels) 
with the decomposition scale is due to the larger dispersion. This feature is 
also the consequence of the smaller number of 
wavelet coefficients at higher decomposition scales.
\begin{table}
\begin{center}
\begin{tabular}{|c|c|c|}
\hline
Scale& $\overline k$ & $\sigma_{\pm}$  \\
\hline
I & 0.04 & 0.15 \\
II & 0.05 & 0.32 \\
III & -0.01 & 0.51 \\
VI & -0.006 & 0.869 \\
\hline
\end{tabular}
\end{center}
\caption{\small\it The mean excess of kurtosis  at four decomposition 
scales computed over the 100 gaussian maps. The $\sigma_{\pm}$ values define 
the confidence intervals for one realisation.}
\label{tab:gng}
\end{table}
\subsection{Partial derivatives }
\begin{table}
\begin{center}
\begin{tabular}{|c|c|c|c|c|c|c|}
\hline
  & Scale& $\overline k_1$ & $\sigma_{\pm}$ & & $\overline k_2$ & $\sigma_{\pm}$\\
\hline
        & I & 0.010 & 0.020  &  & 0.003 & 0.022\\
$\partial/\partial x$ & II & 0.012 & 0.043 & $\partial^2/\partial x\partial y$ & 0.005  & 0.042 \\
   \&        & III & -0.001 & 0.079  &  & 0.006 & 0.072 \\
$\partial/\partial y$ & VI & 0.006 & 0.150 &  & 0.018 & 0.014 \\
\hline
\end{tabular}
\end{center}
\caption{\small\it The mean excess of kurtosis at four decomposition 
scales. $\overline k_1$ is computed using the wavelet coefficients associated 
with 
$\partial/\partial x$ and $\partial/\partial y$. $\overline k_2$ is computed  
using the coefficients of $\partial^2/\partial x\partial y$. The 
$\sigma_{\pm}$ values define the confidence intervals for one realisation.}
\label{tab:momg}
\end{table}
\par
\begin{figure*}
%\epsfxsize=\columnwidth
%\hbox{\epsffile{ds8533f4.eps}}
\resizebox{\hsize}{!}{\includegraphics{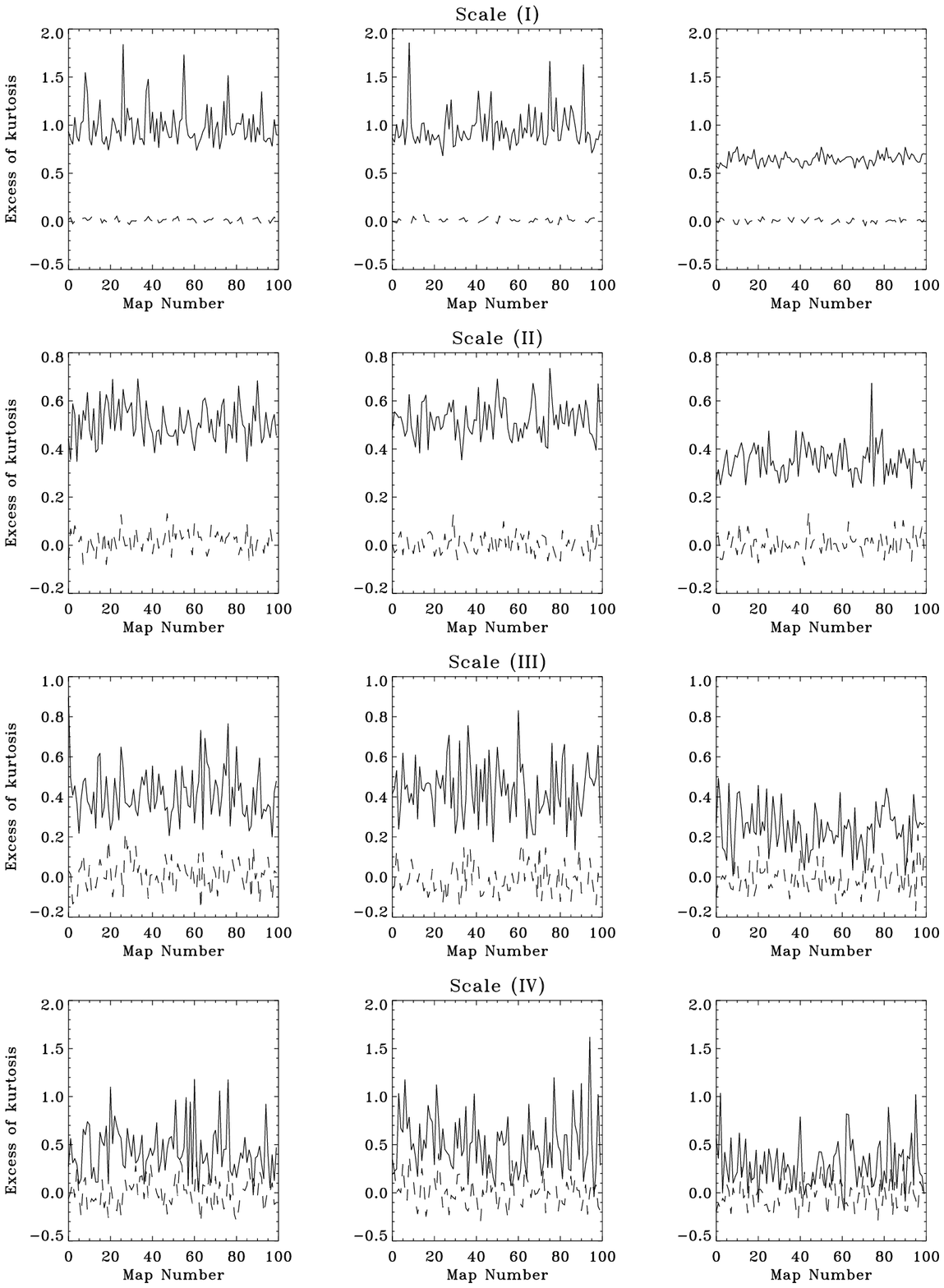}}
\caption{{\small\it Excess of kurtosis computed over the wavelet coefficients
of respectively $\partial/\partial x$ for the left panels, 
$\partial/\partial y$ for the centre panels and $\partial^2/\partial x
\partial y$ for the right panels. The solid line is for the non-gaussian
process. The dashed line is for the gaussian process with same power 
spectrum.}}
\label{fig:mom_bgn}
\end{figure*}
The wavelet
coefficient distributions associated with the first and cross derivatives are
gaussian for the gaussian maps. We thus
compute the normalised excess of kurtosis with respect to a gaussian 
distribution ($k=\mu_4/\mu_2^2-3$). These values are displayed in figure 
\ref{fig:mom_bgn} (dashed line) for the gaussian maps. In this figure and at
each decomposition scale, the first set of 100 values stands 
for the excess of kurtosis of the wavelet coefficients associated with the 
horizontal gradient ($\partial/\partial y$). The second set of 100 values 
represents the same quantity computed for the vertical gradient 
($\partial/\partial x$) and the last one represents the excess of
kurtosis for the wavelet coefficients associated with cross derivative
$\partial^2/\partial x\partial y$ (diagonal gradients).
We note that for the gaussian maps the excess is always centred around zero at 
all the decomposition scales. For the multi-scale gradient, the dispersion 
around the mean $\overline k$  
increases with increasing decomposition scale. In table \ref{tab:momg}, we 
quote the mean  together with the 
confidence intervals
at each scale. The results also show that the values are close to
zero confirming the gaussian nature of the signal.
\par\bigskip
As a result, we conclude that a gaussian signal can be characterised
by the distribution of the multi-scale gradient coefficients and of the 
coefficients associated with $\partial/\partial x$, 
$\partial/\partial y$ and $\partial^2/\partial x\partial y$. In the first
case, the multi-scale gradient coefficient distribution is fitted by a 
Laplace distribution and
the excess of kurtosis is zero. In the second case, the excesses of kurtosis
are gaussian distributed with $\overline k=0$ for the first and the cross
derivatives.\par
We check these two characteristics on other gaussian 
processes with different power spectra. For a white
noise spectrum, we find identical results as in the study case: zero excess of
kurtosis for the multi-scale gradient and the coefficients of the partial 
derivatives.
Since our statistical tests are based on the statistics of the wavelet 
coefficients at each decomposition scale, we expect that a sharp cut off in 
the power spectrum of the gaussian signal will induce a sample variance
problem. We check this behaviour using a gaussian process exhibiting a very
sharp cut off with a shape close to a Heaviside function, at the second 
decomposition scale. This cut off is similar to the cut off expected in the 
CMB power 
spectrum in a standard cold dark matter model. At all the decomposition scales
except the second, we find the expected zero excess of kurtosis for the 
wavelet 
coefficients of gaussian signals. At the second decomposition scale, we find
a non-zero excess of kurtosis for the multi-scale gradient coefficients, as
well as for the coefficients related to the first derivatives, 
which could be misinterpreted for a non-gaussian signature. This non-zero
excess has nothing to do with an intrinsic property of the studied
signal, as the latter is gaussian at all scales. It comes
from the very sharp decrease in power combined with the narrow filter 
associated with the wavelet basis. In fact, the 
contribution of the gaussian process, at this scale, is
sparse. Therefore, it induces a sample variance effect which in turn results 
in a non-zero excess of kurtosis. We tested a wider filtering 
wavelet and found that the excess of kurtosis decreases. 
Nevertheless, a wider filtering wavelet smoothes the non-gaussian signatures 
and reduces the efficiency of our discriminating tests. However, at the second
decomposition scale, the wavelet coefficients associated with  
$\partial^2/\partial x\partial y$ exhibit no excess of kurtosis for
a process with sharp cut off. This indicates that the excess of 
kurtosis computed using the cross derivative coefficients is more reliable in
characterising a gaussian process, and consequently non-gaussianity, 
regardless of the power spectrum. 
\par
We also analysed the sum of gaussian signals. When there is no cut off in the 
power, the excess of kurtosis for
the sum of gaussian signals is zero for both discriminators. By contrast, if 
one of the gaussian
signals presents a cut off at any decomposition scale, we again find a 
non-zero excess of kurtosis at the corresponding scale. 
\section{Application to non-gaussian signals}
We apply our statistical discriminators to detect the non-gaussian signature
of different processes. We first study two sets of non-gaussian maps, 
one constituted of a distribution of top-hat profiles and the other 
constituted of a distribution of gaussian profiles, both having the same power 
spectrum as the gaussian test maps used in the previous section. We also apply 
the statistical test to a combination of gaussian and non-gaussian signals 
with different mixing ratios. 
\subsection{The multi-scale gradient and its distribution}
We compute the multi-scale gradient coefficients (${\mathcal{G}}_L$) using 100 
statistical 
realisations of the non-gaussian process (top-hat profiles). At
the four decomposition scales, we plot the fitted histogram (right panels of 
Figure \ref{fig:hist-gng}). We note that the distribution of the multi-scale 
gradient coefficients also fits a Laplace distribution for small values of 
${\mathcal{G}}_L$. However, there is a significant departure from this 
distribution for higher values. This is exhibited by the larger error bars  
and by the wings of the gradient distribution at large 
${\mathcal{G}}_L$. Figure \ref{fig:hist-gng} (right panels) exhibits the 
non-gaussian 
signatures mostly at the first three decomposition scales. At the fourth,
the lack of coefficients enlarges the error bars but we still marginally 
distinguish the non-gaussian signal.
\begin{table}
\begin{center}
\begin{tabular}{|c|c|c|c||c|c|c|}
\hline
Scale & $k_1$ & $\sigma_{+}$ & $\sigma_{-}$& $k_2$ & $\sigma_+$ & $\sigma_-$\\
\hline
I & 14.50 & 35.38 & 5.06 & 1.5 & 0.51 & 0.27\\
II &  3.87 & 2.68 & 0.77 & 1.9 & 1.43 & 0.48 \\
III & 3.54 & 2.17 & 1.24 & 3.46 & 2.60 & 1.24 \\
VI & 3.57 & 7.13 & 1.68 & 2.88 & 2.55 & 1.82 \\
\hline
\end{tabular}
\end{center}
\caption{\small\it $k_1$ is the median excess of kurtosis, 
at four decomposition 
scales, computed with the 100 non-gaussian maps (top-hat profile). 
$k_2$ is computed with the the 100 non-gaussian maps (gaussian profile).
$\sigma_+$ and $\sigma_{-}$ give the confidence interval for one realisation.}
\label{tab:kng_gp}
\end{table}
\par
In our test case, the process is leptokurtic, that is the non-gaussianity is 
characterised by a positive excess of kurtosis. We quote, in the left panel
of Table \ref{tab:kng_gp} the median excesses of kurtosis  
computed with the multi-scale gradient coefficients of the 100 maps. This 
is a more suitable quantity to characterise a non-gaussian process, than the 
mean $\overline k$, as 
there is an important dispersion of the $k$ values with a clear excess
towards large values. The $\sigma_{\pm}$, which represents the $rms$ excess of 
kurtosis with respect to the median for one realisation, takes naturally into 
account the non symmetric distribution of the multi-scale gradients.
This results in a lower boundary ($\sigma_{-}$) for the confidence interval
smaller that the upper boundary ($\sigma_{+}$). The latter is biased towards 
large values as we are studying a leptokurtic process.
Therefore, the comparison between the values of $k$ and $\sigma_{-}$  
indicates the detectability of non-gaussianity. When
$k-\sigma_{-}$ for one realisation differs from zero by a value of the order 
of, or larger than, $\sigma_{-}$ this suggests that the signal is 
non-gaussian. For the top-hat profiles, there is an obvious excess of
kurtosis at all scales. \par
In order to test the non-gaussian signature arising form different processes
with same power spectra, we analyse a set of 100 non-gaussian maps made of the
superposition of gaussian profiles of different sizes and amplitudes. We 
compute the median value of the excess of kurtosis and the corresponding 
confidence intervals (Tab. \ref{tab:kng_gp}, right panel). We note that $k$
is different from zero at all scales, exhibiting the non-gaussian
nature of the studied process. However, it is smaller than in
the case of the top-hat profiles.
This decrease is due to the superposition of smoother profiles.
\par\bigskip
We now add one representative gaussian map to 100 non-gaussian maps (top-hat
profiles). As the non-gaussian signal is very strongly 
dependent on the studied map, it is necessary to span a 
large set of non-gaussian statistical realisations in order to have a reliable 
statistical
specification of non-gaussianity. The gaussian and non-gaussian
signals were summed 
with different mixing ratios represented by the ratio of their $rms$ 
amplitudes ($R_{rms}=\sigma_{gauss}/\sigma_{non-gauss}$).\\
After wavelet decomposition, we compute the multi-scale gradient coefficients
of the summed maps and derive the normalised median excess of kurtosis with
respect to a Laplace distribution together with the confidence intervals. The 
results are quoted in Table \ref{tab:krms} as a 
function of the mixing ratio $R_{rms}$ and the wavelet decomposition scale. \\
For $R_{rms}=1$, the excess of kurtosis is larger than that of the gaussian
test map and it is smaller than that of the purely non-gaussian signal. The 
summation of
the two processes has therefore, as expected, smoothed the gradients 
and diluted 
the non-gaussian signal. For a non-gaussian signal half that of the
gaussian signal, only the first three scales indicate an excess of kurtosis
different from the gaussian one. For the ratio $R_{rms}=3$, only the first
scale has an excess marginally different from the gaussian signal. For larger
ratios, the non-gaussian signal is quite blurred. 
\begin{table}
\begin{center}
\begin{tabular}{|c|c|c|c|c|}
\hline
Ratio      & Scale & $k$ & $\sigma_+$ & $\sigma_-$\\
\hline
$R_{rms}=1$&  I & 6.64 & 12.18 & 2.36\\
           & II & 1.13 & 0.80 & 0.41 \\
           & III & 0.78 & 1.42 & 0.57 \\
           & VI & 0.62 & 2.81 & 0.82 \\
\hline
$R_{rms}=2$     & I & 1.64 & 3.42 & 0.56 \\
           & II & 0.24  & 0.39 & 0.23 \\
           & III & 0.06 & 0.77 & 0.34 \\
           & VI & 0.23 & 1.08 & 0.58 \\
\hline
$R_{rms}=3$     & I & 0.36 & 0.44 & 0.11 \\
           & II & 0.09  & 0.16 & 0.09 \\
           & III & 0.04 & 0.26 & 0.21 \\
           & VI & 0.26 & 0.37 & 0.36 \\
\hline
$R_{rms}=4$    & I & 0.33 & 0.15 & 0.06 \\
           & II & 0.09  & 0.08 & 0.06 \\
           & III & 0.05 & 0.13 & 0.13 \\
           & VI & 0.27 & 0.19 & 0.23 \\
\hline

\end{tabular}
\end{center}
\caption{\small\it The median excess of kurtosis $k$, at four decomposition 
scales,computed over the 100 non-gaussian maps added to the gaussian map as
a function of the mixing ratios $R_{rms}$ and the decomposition scales. The 
$\sigma_+$ and $\sigma_-$ values represent respectively the upper
and lower boundaries of the confidence interval for one realisation. They are 
the $rms$ values with respect to the median excess of kurtosis.}
\label{tab:krms}
\end{table}

\subsection{Partial derivatives}
\begin{table}
\begin{center}
\begin{tabular}{|c|c|c|c|c|c|c|c|}
\hline
& Scale & $k_1$ & $\sigma_+$ & $\sigma_-$ & & $k_2$ & $\sigma_\pm$\\
\hline
&I & 0.92 & 0.28  & 0.09 & & 0.65 & 0.05 \\
$\partial/\partial x$ & II & 0.51 & 0.08 & 0.07 &$\partial^2/\partial x\partial y$ & 0.35  & 0.06 \\
  \&      & III & 0.41 & 0.15 & 0.11 & & 0.24 & 0.11 \\
$\partial/\partial y$  & VI & 0.43 & 0.34 & 0.21& & 0.28 & 0.21\\
\hline
\end{tabular}
\end{center}
\caption{\small\it The median excess of kurtosis, at four decomposition 
scales. $k_1$ is computed using the wavelet coefficients associated with 
$\partial/\partial x$ and $\partial/\partial y$. $k_2$ is given for  
$\partial^2/\partial x
\partial y$. The 100 non-gaussian maps (top-hat profile) have the same 
power spectrum as the 
gaussian test maps. The $\sigma_+$ and $\sigma_-$ values are the $rms$ values 
for one realisation with respect to the median excess of kurtosis.}
\label{tab:momng}
\end{table}
For the 100 non-gaussian maps (top-hat profile) with the same power spectrum 
as the gaussian test 
maps, we compute the normalised excess of kurtosis, with respect to a gaussian,
of the wavelet coefficients associated 
with $\partial/\partial x$, $\partial/\partial y$ and
$\partial^2/\partial x\partial y$. As for the multi-scale gradient,
we derive the median excess of kurtosis and the upper and lower
boundaries of the confidence intervals. The results, given in
Table \ref{tab:momng}, show non-zero excesses of kurtosis for the first and
cross derivatives at all decomposition scales. \par
In Figure \ref{fig:mom_bgn}, the solid line represents the values of the
excess of kurtosis of each non-gaussian realisation. The dashed line 
represents the same quantity for the gaussian test maps. We first note the 
overall shift of 
the values towards non-zero positive values (leptokurtic signal), with some 
very
large values compared to the median. A second characteristic worth noting is 
the difference in amplitudes between, on the one hand, the excess of kurtosis 
of the coefficients
associated with $\partial^2/\partial x\partial y$ and, on the other hand, 
those associated with $\partial/\partial x$ and $\partial/\partial y$. 
The former are indeed smaller. As the excess 
of kurtosis of the first derivative coefficients is of the same order, we 
compute one median $k$ over $\partial/\partial x$ and $\partial/\partial y$
coefficients, and compare it to the excess of kurtosis of the cross derivative.
At the first two decomposition  scales there is an important and 
noticeable difference between 
the two sets of values $\partial/\partial x$ and $\partial/\partial y$,
and $\partial^2/\partial x\partial y$. At the third 
and fourth decomposition scales, the difference decreases but is still present.
\begin{table}
\begin{center}
\begin{tabular}{|c|c|c|c|c|c|c|c|c|}
\hline
Scale & & $k_1$ & $\sigma_+$ & $\sigma_-$ & & $k_2$ & $\sigma_{\pm}$ \\
\hline
I & &0.19 & 0.03 & 0.02 & & 0.10 & 0.03 \\
II & $\partial/\partial x$ & 0.21 & 0.07 & 0.06 & & 0.13 & 0.05 \\
III & \& & 0.35 & 0.14 & 0.13 & $\partial^2/\partial x\partial y$ & 0.17 & 0.10 \\
VI & $\partial/\partial y$ & 0.24 & 0.24 & 0.18 & & 0.21 & 0.24 \\
\hline
\end{tabular}
\end{center}
\caption{\small\it The median excess of kurtosis, at four decomposition 
scales, computed over the 100 non-gaussian maps
(gaussian profile). $k_1$ is the median excess computed with 
the coefficients
associated with the vertical and horizontal gradients, and $k_2$ is given 
for the cross derivative. The $\sigma$ values are the boundaries
of the confidence interval for one statistical realisation.}
\label{tab:mbgng}
\end{table}
\par
For the non-gaussian process made of the superposition of gaussian profiles,
we compute the median excess of kurtosis associated with the wavelet 
coefficients of the first and cross derivatives (Tab. \ref{tab:mbgng}). As for
the multi-scale gradient coefficients, we find that the excess of kurtosis
is smaller for this type of non-gaussian maps but it is still significantly 
different from zero at all scales except the fourth.
\begin{table*}
\begin{center}
\begin{tabular}{|c|c|c|c|c|c|c|c|c|}
\hline
$R_{rms}$ & Scale & & $k_1$ & $\sigma_+$ & $\sigma_-$ & & $k_2$ & $\sigma_{\pm}$ \\
\hline
1 & I & &0.44 & 0.14 & 0.06 & & 0.25 & 0.03 \\
& II & $\partial/\partial x$ & 0.15 & 0.06 & 0.05 & & 0.09 & 0.03 \\
& III & \& & 0.09 & 0.11 & 0.08 & $\partial^2/\partial x\partial y$ & 0.03 & 0.08 \\
& VI & $\partial/\partial y$ & 0.10 & 0.22 & 0.16 & & -0.06 & 0.08 \\
\hline
2 & I & &0.13 & 0.06 & 0.03 & & 0.07 & 0.02 \\
& II & $\partial/\partial x$ & 0.04 & 0.04 & 0.04 & & -0.002 & 0.030 \\
& III & \& & -0.002 & 0.090 & 0.068 & $\partial^2/\partial x\partial y$ & -0.07 & 0.06 \\
& VI & $\partial/\partial y$ &0.03 & 0.13 & 0.13 & & -0.20 & 0.08 \\
\hline
3 & I & &0.05 & 0.04 & 0.03 & & 0.03 & 0.02 \\
& II & $\partial/\partial x$ & 0.02  & 0.04 & 0.04 & & -0.02 & 0.02 \\
& III & \& & -0.007 & 0.078 & 0.071 & $\partial^2/\partial x\partial y$ & -0.10 & 0.04 \\
& VI & $\partial/\partial y$ & 0.01 & 0.11 & 0.11 & & -0.26 & 0.06 \\
\hline
4 & I & &0.03 & 0.04 & 0.03 & & 0.03 & 0.01 \\
& II & $\partial/\partial x$ & 0.02 & 0.04 & 0.04 & & -0.02 & 0.02 \\
& III & \& & -0.006 & 0.072 & 0.074 & $\partial^2/\partial x\partial y$ & -0.11 & 0.03 \\
& VI & $\partial/\partial y$ & 0.003 & 0.097 & 0.078 & & -0.27 & 0.04 \\
\hline
\end{tabular}
\end{center}
\caption{\small\it The median excess of kurtosis, at four decomposition 
scales, for 100 non-gaussian maps (top-hat profile) added 
to one gaussian map (with 
same power spectrum) as a function of the mixing ratio $R_{rms}$. $k_1$ is 
computed with the coefficients associated with $\partial/\partial x$ 
and $\partial/\partial y$, and $k_2$ is given for  
$\partial^2/\partial x\partial y$. The $\sigma$ values are the boundaries
of the confidence interval for one statistical realisation.}
\label{tab:moms}
\end{table*}
\begin{figure*}
%\epsfxsize=\columnwidth
%\hbox{\epsffile{ds8533f5.eps}}
\resizebox{\hsize}{!}{\includegraphics{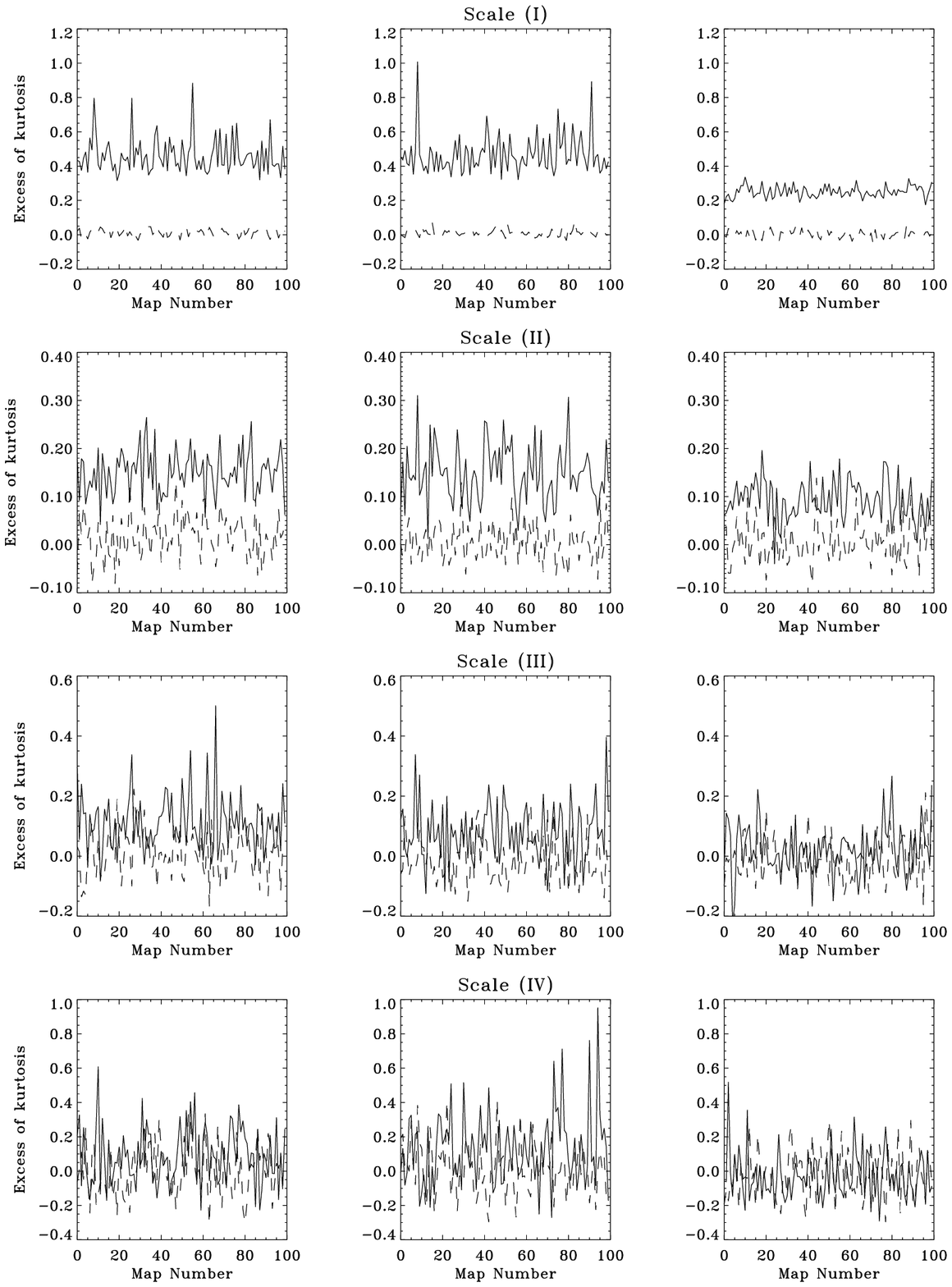}}
\caption{{\small\it Excess of kurtosis computed with the wavelet coefficients
of $\partial/\partial x$ and 
$\partial/\partial y$ for respectively the left and centre panels. 
The right panels are associated with the coefficients of 
$\partial^2/\partial x\partial y$. The gaussian and non-gaussian processes
have same power spectra and have been summed with a mixing ratio $R_{rms}=1$.
 The dashed line represents the gaussian process alone and the 
solid line represents the sum of the two processes.}}
\label{fig:mom_bng}
\end{figure*}
\par\bigskip 
We analyse the sum of a representative gaussian map and the set of 100
non-gaussian maps (top-hat profile). 
The sum of the two processes has been performed again with different 
mixing ratios. The results we obtain are given in 
Table \ref{tab:moms}. An accompanying figure (Fig. \ref{fig:mom_bng}) 
illustrates the corresponding results for a mixing ratio $R_{rms}=1$ (solid 
line).
In this figure, the dashed line represents the gaussian process alone. 
For $R_{rms}=1$, we find that non-gaussianity is detected at
all decomposition scales for both first and cross derivative
coefficients. For a mixing ratio $R_{rms}=2$, we
observe a significant excess of kurtosis only at the first decomposition 
scale. For $R_{rms}\ge 3$, the excess becomes marginal for both 
$\partial/\partial x$ and $\partial/\partial y$ coefficients at the first
decomposition scale, all other scales showing no departure from gaussianity. 
The same tendency is noted for the coefficients associated with
$\partial^2/\partial x\partial y$. 
\section{Detection strategy of non-gaussianity} \label{sec:metod}
We have characterised the gaussian signal through the excess of kurtosis of 
the multi-scale gradient and partial (first and cross) derivative coefficients 
using processes with different power spectra. When the power spectrum of the 
process exhibits a sharp cut off in one of the wavelet decomposition windows
we find that the excess of kurtosis, associated with the coefficients of the
first derivatives, and consequently of the multi-scale gradient, is non-zero
at the filtering level of the cut off. On the contrary, the excess of kurtosis
computed with the cross derivative coefficients is zero.  
\par
Accordingly, we propose a detection strategy to test for non-gaussianity. 
We compare a set of maps of the ``real'' observed sky to a set of gaussian 
realisations having the power spectrum of the ``real sky''. Our
proposed method overcomes the problems arising from eventual cut off in the 
power spectrum of the 
studied process, and the consequent possible misinterpretations on the 
statistical signature. It also constitutes the most general approach to 
exhibit the statistical nature
(gaussian or not) of a signal and quantify its detectability through our 
statistical tests. Our detection strategy of the non-gaussianity
is based on the following steps: \par
$\bullet$ using observed maps of the ``real sky'' we compute the angular power 
spectrum of the signal, regardless of its statistical nature. \par
$\bullet$ We simulate gaussian synthetic realisations  
of a process having the power spectrum of the ``real'' process. On the 
obtained gaussian test maps filtered with the wavelet function, we compute the 
excess of kurtosis for the multi-scale gradient and derivative coefficients. 
This analysis allows us to
characterise completely the gaussian maps, naturally taking into account 
eventual sample variance effects due to cut off at any scale. \par
$\bullet$ For the set of observed maps of the ``real sky'', we compute the 
excesses of kurtosis associated with the multi-scale gradient, and the 
derivative coefficients. \par
$\bullet$ Assuming that the realisations (maps) are independent, each value of 
the excess of kurtosis has a probability of $1/N$, where $N$ is the number of
maps. 
Using the computed excesses of kurtosis of both the gaussian and non-gaussian 
realisations, we deduce the probability distribution function (PDF) of the 
excess of kurtosis, for the multi-scale gradient coefficients and for the 
coefficients related to the derivatives. \par
$\bullet$ The last step consists of quantifying the detectability of the 
non-gaussian signature. That is to compare the PDF of the gaussian process to
the PDF of the ``real sky''. 
In practice this can be done by computing at each decomposition scale the 
probability 
that the median excess of kurtosis of the non-gaussian maps belongs to the 
PDF of the synthetic gaussian counterparts. It is the probability that a 
random 
variable is greater or equal to the real median $k$. We take $k$ which is the 
asymptotic value given by the central limit theorem. Another way of comparing 
the two PDFs is to use the Kolmogorov-Smirnov (K-S) test (Press et al. 1992) 
which gives the probability
for two distributions to be identical. This test for non-gaussianity is more 
global than the previous test because it is sensitive to the shift in the PDFs,
especially the median value, and to the spread of the distributions. This 
property makes it more sensitive to non-gaussianity especially in the case 
where we only have a small number of observed maps.
\par
\begin{figure}
\resizebox{\hsize}{!}{\includegraphics{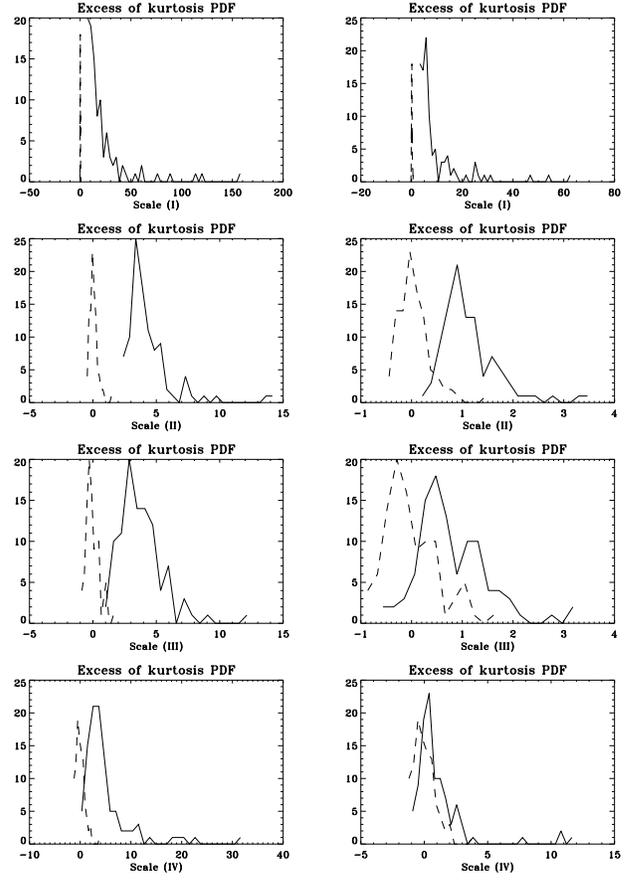}}
\caption{{\small\it Probability distribution functions of the excess of 
kurtosis, as percentages, computed with the multi-scale gradient coefficients. 
In the two panels, the dashed line represents PDF of 
the gaussian test maps. In the left panels, the solid line represents the
non-gaussian signal alone (top-hat profiles). In the right panels, the solid 
line represents the sum of the gaussian and non-gaussian processes with 
$R_{rms}=1$.}}
\label{fig:pdfk}
\end{figure}

\begin{figure}
\resizebox{\hsize}{!}{\includegraphics{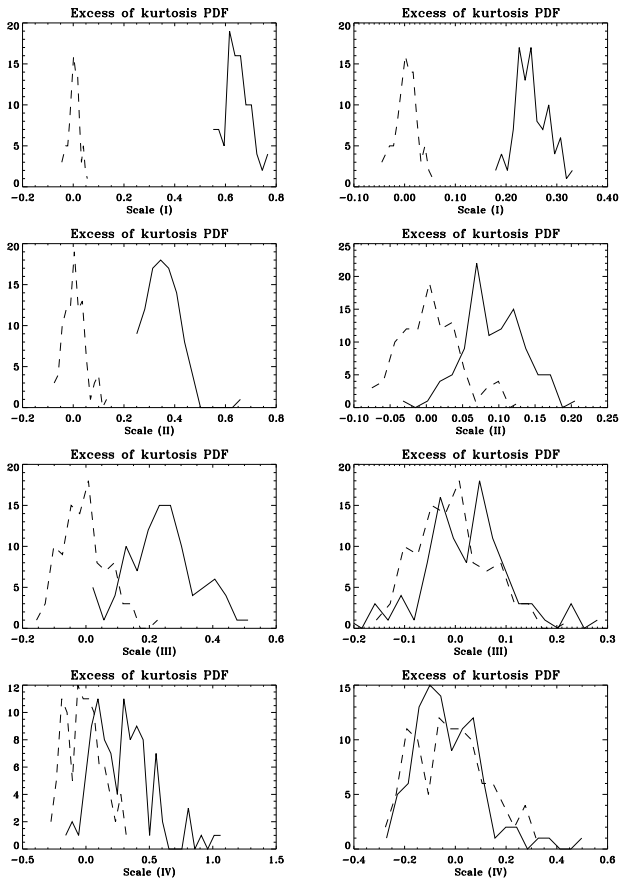}}
\caption{{\small\it Probability distribution functions of the excess of 
kurtosis, as percentages, computed with the coefficients associated with the 
cross derivative.
In the two panels, the dashed line represents PDF of 
the gaussian test maps. In the left panels, the solid line represents the
non-gaussian signal alone (top-hat profiles). In the right panels, the solid 
line represents the sum of the gaussian and non-gaussian processes with 
$R_{rms}=1$.}}
\label{fig:pdfc}
\end{figure}

\bigskip
We apply our detection strategy of non-gaussianity to the non-gaussian 
test maps constituted of the top-hat and gaussian profiles. For illustrative
purposes, we give the results of the multi-scale gradient coefficient only. At 
the first three decomposition scales and for both sets of maps, we find 
that the probability for the signal to be non-gaussian is 100\% 
using the probability of the measured $k$ to belong to the gaussian PDF. At the
fourth scale, the probability is 99.99\% and 99.95\% for respectively the 
top-hat and gaussian profile distribution. The K-S test gives a 100\% 
probability of detecting non-gaussianity. The detectability of the non-gaussian
signature, for the sum of the gaussian and non-gaussian (top-hat profile) maps 
with mixing ratio
$R_{rms}=1$, is 100\% at the first decomposition scale. It is 99.96\% and 
93.4\% at the second and third scale, and 76.43\% at the fourth scale. For 
$R_{rms}=2$, the first scale is still perfectly 
non-gaussian, and only the second scale is detected with a probability of
72.4\%. The K-S test gives more or less the same results for both mixing 
ratios. The results are illustrated in Figure \ref{fig:pdfk} (for the
multi-scale gradient coefficients) and in Figure \ref{fig:pdfc} (for the cross 
derivative coefficients). In these plots, the solid line represents the 
PDF of the excess of kurtosis for the non-gaussian measured signal.
The dashed line represents the PDF of the 
synthetic gaussian maps with same power spectrum. In both  
figures the left panels are for the non-gaussian signal 
alone, whereas the right panels are for the sum with a mixing ratio of 
one.

\section{Discussion and Conclusions}
In the present work we develop two statistical discriminators to test for
non-gaussianity. 
To do so, we study the statistical properties of the coefficients in a four 
level dyadic wavelet decomposition.\par
Our first discriminator uses the amplitude of the multi-scale gradient
coefficients. It is based on the computation of their excesses of kurtosis with
respect to a Laplace distribution function. The second test relies on the 
computation of the excesses of kurtosis for the first and
cross derivative coefficients. It can itself be divided into one specific 
test using the first derivatives and the other the cross derivative. 
For both discriminators (multi-scale gradient and partial derivatives), the 
gaussian signature is characterised by a zero excess of kurtosis.
We check this property for several gaussian processes with different power 
spectra and for
a signal made of the sum of gaussian signals. Given this property for 
gaussianity, the departure form a zero value
of the excess of kurtosis indicates the non-gaussian signature. \par
In order to overcome peculiar features in the power spectrum (e.g. sharp
cut offs) at any wavelet decomposition scale, which could be misinterpreted for
a non-gaussian signature, we propose the following detection strategy. We
simulate synthetic gaussian maps with the same power spectrum as the 
non-gaussian studied
signal. We compute the excess of kurtosis for the two discriminators, and for 
both gaussian and non-gaussian maps. We derive the PDF
in each case. Then, we quantify the detectability of non-gaussianity 
by estimating the
probability that the median excess of kurtosis of the non-gaussian signal 
belongs to 
the PDF of the gaussian counterpart, and by 
applying the K-S test to discriminate the gaussian and the ``real'' PDFs. 
We apply our detection strategy to the test maps of non-gaussian signals 
alone, and to the sum of gaussian and non-gaussian signals. In the first 
case, we show that the non-gaussian signature emerges clearly at all 
scales. In the second case, the detection depends on the mixing ratio
($\sigma_{gauss}/\sigma_{non-gauss}$). Down to a mixing ratio of about 3, 
which is about 10 in term of power, we detect the non-gaussian signature.
\par
In parallel to our work, Hobson et al. (1998) have used the wavelet coefficients
to distinguish the non-gaussianity due to the Kaiser-Stebbins effect 
(Bouchet et al 1988; Stebbins 1988) of cosmic strings. They used the 
cumulants of the wavelet coefficients up to the fourth order 
(Ferreira et al 1997), in a pyramidal decomposition. As mentioned in section 
\ref{sec:wdec} and in Hobson et al. (1998), the pyramidal decomposition induces 
a scale mixing. Therefore, it does not take advantage of the possible spatial 
correlations of the signal. Furthermore, it gives smaller numbers of 
coefficients within each sub-band for the analysis. 
We instead use the dyadic decomposition to avoid these two weaknesses, as in
Aghanim et al. (1998). \par
In our study, we use weakly non-gaussian simulated maps (small kurtosis). Such 
a weak non-gaussian signature, by contrast with the 
Kaiser-Stebbins effect, and with the point-like or peaked profiles, is 
detected using our
statistical discriminators. Using the bi-orthogonal wavelet transform, we 
succeed in emphasising the
non-gaussianity, by the means of the statistics of the wavelet coefficient
distributions. This detection is also possible
using other bi-orthogonal wavelet bases, but their efficiency
is lower at larger scales. Consequently, the choice of the wavelet basis 
depends also on the characteristics of the non-gaussian signal one wants to
emphasise. However we believe that the wavelet basis we choose represents
an optimal compromise for a large variety of non-gaussian features.
\begin{acknowledgements}
The authors wish to thank the referee A. Heavens for
his comments that improved the paper. We also thank F.R. Bouchet, 
P. Ferreira and J.-L. Puget for valuable
discussions and comments and A. Jones for his careful reading.
\end{acknowledgements}

\end{document}